\documentclass[prl,twocolumn,superscriptaddress,showpacs]{revtex4-1} 
\usepackage{epsfig}
\usepackage{graphics}
\usepackage{float}
\usepackage{amsmath}
\usepackage{xcolor,colortbl}
\usepackage[utf8]{inputenc}
\usepackage[T1]{fontenc}
\usepackage{lmodern}
\usepackage{soul}
\usepackage{todonotes}

\definecolor{Gray}{rgb}{0.72,0.72,0.98}
\definecolor{LightCyan1}{rgb}{0.83,0.83,0.98}
\definecolor{LightCyan2}{rgb}{0.91,0.92,1}

\begin{document}

\newcommand{\upn}{Departamento de Ciencias, Universidad Privada del Norte, Lima 15434, Peru.}

\newcommand{\cnr} {Istituto di Struttura della Materia of the National Research Council, Via Salaria Km 29.3,I-00016 Monterotondo Stazione, Italy.}
\newcommand{\etsf} {European Theoretical Spectroscopy Facilities (ETSF)}
\newcommand{\ift} {Instituto de F\'{\i}sica Te\'{o}rica, Universidade Estadual Paulista (UNESP), Rua Dr. Bento T. Ferraz, 271, S\~{a}o Paulo, SP 01140-070, Brazil.}
\newcommand{\unk} {Department of Materials Science and Engineering, University of California Davis, CA, 95616 USA.}

\title{Efficient hot carrier dynamics in near-infrared photocatalytic metals}

\author{Cesar E. P. Villegas}
\affiliation{\upn}
\author{Marina S. Leite}
\affiliation{\unk}
\author{Andrea Marini}
\affiliation{\cnr}
\author{A. R. Rocha}
\affiliation{\ift}

\date{\today}

\begin{abstract}
Photoexcited metals can produce highly-energetic hot carriers whose controlled generation and extraction is a promising avenue for technological applications. While hot carrier dynamics in Au-group metals have been widely investigated, a microscopic description of the dynamics of photoexcited carriers in the mid-infrared and near-infrared Pt-group metals range is still scarce. Since these materials are widely used in catalysis and, more recently, in plasmonic catalysis, their microscopic carrier dynamics characterization is crucial. We employ \emph{ab initio} many-body perturbation theory to investigate the hot carrier generation, relaxation times, and mean free path in bulk Pd and Pt. We show that the direct optical transitions of photoexcited carriers in this metals are mainly generated in the near-infrared range. We also find that the electron-phonon mass enhancement parameter for Pt is 16 $\%$ higher than Pd, a result that help explains several experimental results showing diverse trends.
Moreover, we predict that Pd (Pt) hot electrons possess total relaxation times of up to 35 fs (24 fs), taking place at approximately 0.5 eV (1.0 eV) above the Fermi energy. Finally, an efficient hot electron generation and extraction can be achieved in nanofilms of Pd (110) and Pd (100) when subject to excitation energies ranging from 0.4 to 1.6 eV.
\end{abstract} 

\maketitle
\section{Introduction}
\begin{figure*}[ht]
\centering
\includegraphics[width=1.7\columnwidth,height=7.5cm]{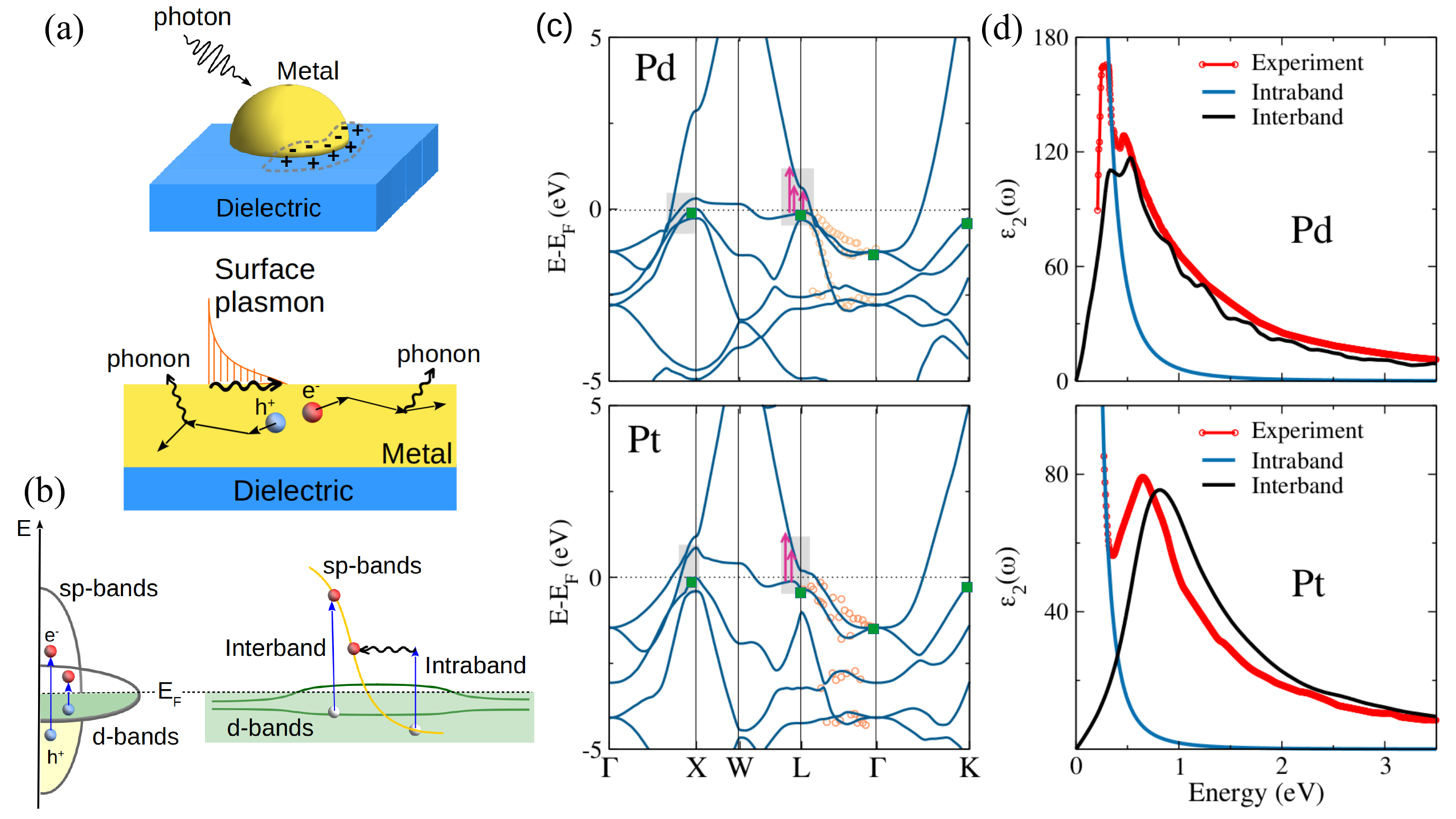}
\caption{\small{ a) Schematic representation of the formation of a surface plasmon and its subsequent decay into hot carriers (hot electron and hot hole). The inelastic phonon scattering processes of hot carriers and their corresponding relaxation path is depicted by the arrows. b) Simplified representation of density of states for near- infrared plasmonic metals. c) Quasiparticle electronic bandstructure for Pd and Pt (blue lines) obtained within the real-axis GW approximation compared with angle-resolved ultraviolet photoemission measurements (orange dotted line) \cite{himpsel1978,eyers1984}. The gray shaded regions indicate the crystallographic directions where the most relevant direct optical transitions occur. d) Comparison of the calculated interband and intraband absorption with experimental measurements for Pd \cite{lafait1978} and Pt \cite{weaver1974}.}}\label{fig1} 
\end{figure*}

Noble metals belong to a class of materials whose physical properties have been extensively studied, both theoretically \cite{xu2017,choi2006,marini2002,tang2011,wilson2020,gall2016} and experimentally \cite{werner2019,olmon2012,suemoto2021,garcia2019}. This has lead to a wide range of applications in photovoltaics \cite{linic2011,atwater2010}, plasmonics \cite{jiang2017,gong2016}, and most recently in an emergent area called plasmonic catalysis \cite{aslam2017,aslam2018,cortes2020}. Some of these metals' outstanding properties stem from the formation of large surface areas, high electron densities and strong interaction of light with plasmons. This enables the absorption of vast amounts of electromagnetic radiation near the metallic surface, a phenomenon known as surface plasmon resonance  \cite{yu2019,barnes2003}. Thus, photoexcitated metals can produce highly-energetic hot carriers, via non-radiative surface plasmon decay \cite{sonnichsen2002,maier2007}, whose subsequent dynamics strongly depends on inelastic electron-electron ($e$-$e$), electron-phonon ($e$-$ph$), and defect scattering processes \cite{fatti2000,brown2016}, as schematically shown in Figure \ref{fig1}a.

As demonstrated in recent experiments, the efficient generation and extraction of photoexcitated carriers, which ultimately depends on their hot carrier relaxation time and mean free path, may be used in a number of novel applications \cite{clavero2014,brongersma2015,ng2016}. Therefore, having a precise knowledge of -- and perhaps control over -- the microscopic processes that hot carriers undergo is of utmost importance for efficient device development. 

While hot carrier dynamics in Au-group metals (Au,Ag,Cu) covering the ultraviolet and visible spectral range have been widely studied due to their highly-energetic efficient generation via interband excitations,\cite{gall2016,bernardi2015,brown2016} a microscopic understanding of the carrier relaxation processes in Pt-group metals (Pt,Pd,Ni) have received little attention \cite{suemoto2021}. These carriers are generated in the mid-infrared and near-infrared range, and 
subsequent engineering of the electronic and optical properties by alloying them with other noble metals has emerged as a route for the efficient generation of hot carriers as this procedure brings the positions of the $d$-band closer to the Fermi level, E$_F$ \cite{gong2020,gong2018,dias2019,mcclure2019}. Consequently it reduces the interband energy threshold, permitting the efficient generation of long-lived hot carriers in the infrared and near-infrared region \cite{stofela2020,krayer2020}.

Accurate theoretical descriptions of catalytic metals, including their electronic \cite{ostlin2016,ladstadter2004,windmiller1969,andersen1970}, vibrational \cite{dalcorso2007,shen2016,savrasov1996,xu2020}, and optical \cite{weaver1974,glantschnig2010,lafait1978} properties abound. Nonetheless, \emph{ab initio} calculations addressing the scattering relaxation rates and mean free paths are still scarce. Indeed, the lack of predictive studies concerning the microscopic e-ph scattering mechanism become especially critical, in light of the available experimental measurements indicating that Pd and Pt present large mass enhancement parameters, $\lambda(E_{F})$, which play an important role in many physical phenomena such as electronic heat capacity and superconductivity. This is even more striking, since these metals do not exhibit superconductivity despite the large value of $\lambda(E_{F})$-- ranging from 0.6 to 0.7-- which arises from the contribution of the electron-phonon and electron-paramagnon interactions. The later (the spin fluctuations) has been suggested to be the responsible for the suppression of the superconductive state \cite{allen1987,knapp1972}. In addition, there is a long standing debate over the hierarchy of enhancement factors between Pt and Pd. Available specific heat measurements suggest that $\lambda_{Pd}$ > $\lambda_{Pt}$, which implies that the electron-phonon lifetime at the Fermi energy for Pt is greater than that of Pd \cite{knapp1972,andersen1970}. This conclusion is, however, contrary to what is estimated from semi-empirical electron-phonon lifetimes obtained from resistivity measurements, in which the lifetimes of Pd are slightly higher compared Pt.

We set out to elucidate this set of interesting phenomena, all of which, closely related to the inelastic e-ph and e-e scattering rates. We do so by carrying out \emph{ab initio} many-body perturbation theory \cite{Onida2002} to study the hot carrier generation, carrier lifetimes, and carrier transport in bulk Pd and Pt. Our results show that the photoexcited carriers in these metals are mainly generated in the near-infrared range by direct optical transitions. We also found that the electron-phonon mass enhancement parameter for Pt is $16 \%$ higher than Pd, a result that corroborate recent experimental trends obtained by resistivity \cite{wilson2020} and ultrafast luminescence measurements \cite{suemoto2021}. This is due to a higher than previously obtained density of states for Pt at the Fermi level. Finally, we show that thin films of Pd (110) and Pd (100) subject to excitation energies ranging from 0.4 to 1.6 eV can be used for efficient hot carrier generation and extraction as a consequence of its 30 nm mean free path, comparable to that of noble metals.
\section{Results and discussion}
\begin{figure*}[t]
\centering
\includegraphics[width=1.8\columnwidth,height=8.5cm]{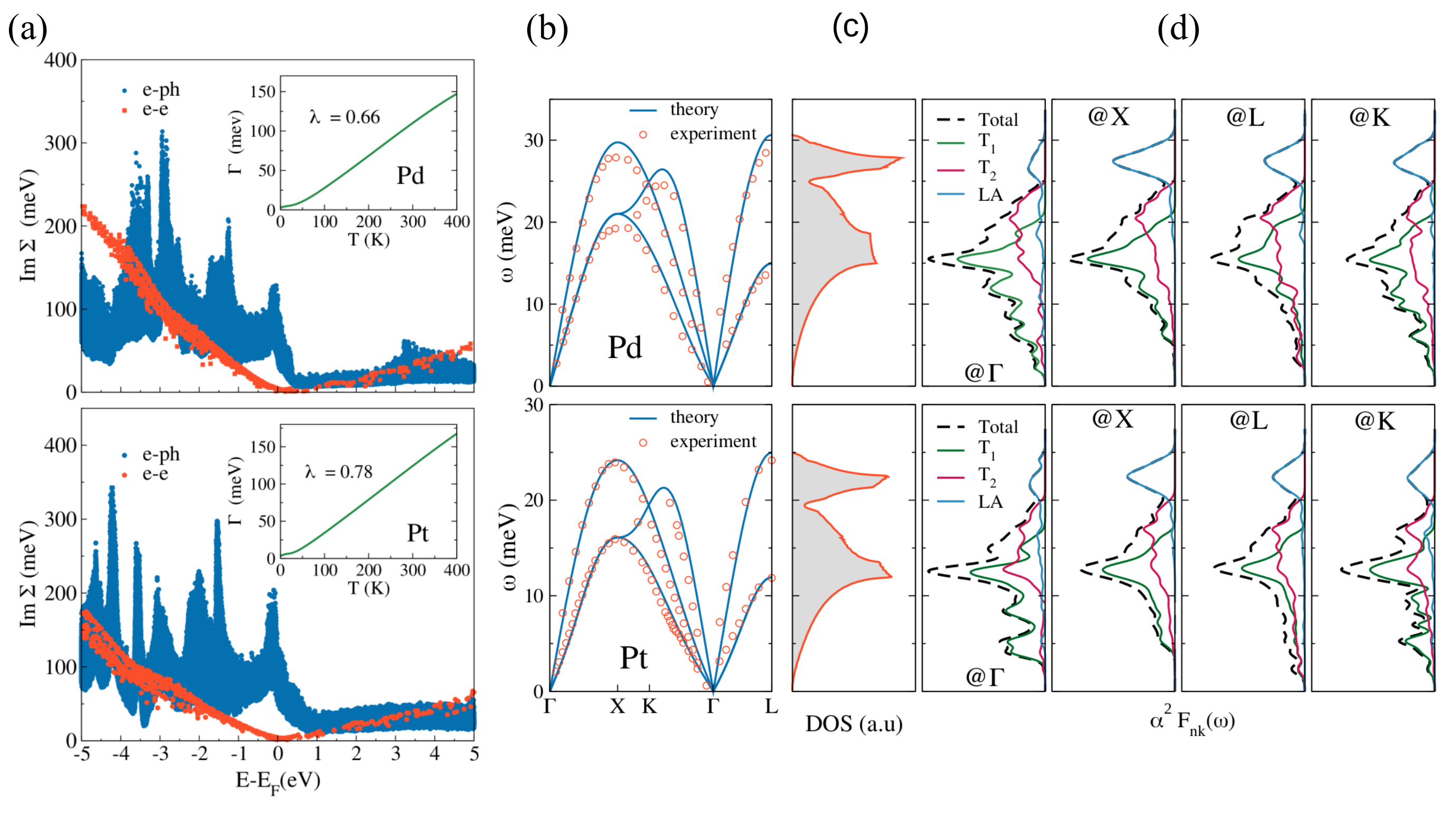}
\caption{\small{a) Imaginary part of the $e$-$e$ and $e$-$ph$ self-energies for Pd and Pt. The electron-phonon rates were computed at room temperature. The insets show the temperature dependence of the electron-phonon linewidth at Fermi energy. b) electron-phonon spectral function $\alpha^{2}_{tr} F(\omega)$ (solid lines) and phonon density of states (shaded curve). The phonon DOS curve is normalized to enclose the same area as the spectral function. c) comparison of the calculated phonon dispersion and experimental measurements for Pd \cite{miiller1968} and Pt \cite{dutton1972}. Right panels shows the generalized Eliashberg spectral function at the highest occupied energy states for the high-symmetry points $G$, $X$, $L$ and $K$.}}\label{fig2} 
\end{figure*}
In metals, the dielectric response function, which gives the photon absorption probability, is composed by two contributions. First i) the intraband one, $\epsilon_{2}^{intra}(\omega)$: indirect electronic excitations between partially filled $sp$ bands, that require a transfer of momentum provided by either phonons or plasmons. This typically occurrs in the infrared energy range. Second ii) due to the interband transitions $\epsilon_{2}^{inter}(\omega)$: direct electronic excitations from occupied $d$ bands to unoccupied $sp$ states, which causes a strong absorption peak. These two contributions are schematically represented in Figure \ref{fig1}-b. Given that in Pd and Pt, the most intense peaks of the dielectric function are located in the infrared and near-infrared spectral range, the intraband and interband contributions may overlap. Thus, separetely analyzing each contributions is a requirement.
In order to understand the microscopic mechanism of photocarrier generation, governed by the electronic structure, in Figure \ref{fig1}c we present the quasiparticle GW bandstructure, fully including spin-orbit interactions, which are compared with experimental data from angle-resolved ultraviolet photoelectron spectroscopy along the $\Gamma L$ path. Clearly, our predicted  QP energies show excellent agreement with experimentally available data in the vicinity of the $L$ point, close to the Fermi energy, whose electronic states are relevant for the carrier photo-excitation in the near-infrared energy window.
As shown, the Fermi level is intersected by $d$ bands narrowing the energy window for allowed $d$-to-$sp$ transitions, when compared with gold group metals (see supplementary information).
A careful analysis of the oscillator strength for both metals reveals that these optical transitions occur around the high symmetry points $L$ and $X$, which are schematically represented by the shaded regions in Figure \ref{fig1}c.  In Figure \ref{fig1}d the interband contribution of the dielectric function is obtained fully $\emph{ab-initio}$ while the intraband one is computed by substituting the estimated experimental electron lifetime and plasma frequency for Pd \cite{lafait1978} and Pt \cite{weaver1974}, into the Drude model. Our quasiparticle optical absorption for Pd shows two intense peaks located at 0.33 and 0.53 eV, which are in excellent agreement with the experimental measurements obtained by Lafait \emph{et al} \cite{lafait1978}. Accordingly, the calculated absorption spectrum of Pt presents an intense peak at 0.76 eV which is less than 100 meV larger than the one 0.67 eV found in experimental measurements \cite{weaver1974}.
Furthermore, notice that the Pd intraband contribution strongly overlaps with the interband one in the mid-infrared region. This result clarify an old debate since it was not possible to unambiguously distinguish in experiments the intraband and interband contributions in Pd \cite{lafait1978,weaver1974,vargas2019}, a task that became more challenging due to the lack of accurate band structure calculations. In contrast, the intraband contribution of Pt weakly overlaps with the interband one. It should be noted that the weak (strong) overlap between intraband and interband contributions for Pt (Pd) is directly ascribed to the position of $d$ states (around the high symmetry point $L$) with respect to the Fermi energy. 
\begin{figure*}[t]
\centering
\includegraphics[width=1.50\columnwidth,height=8.5cm]{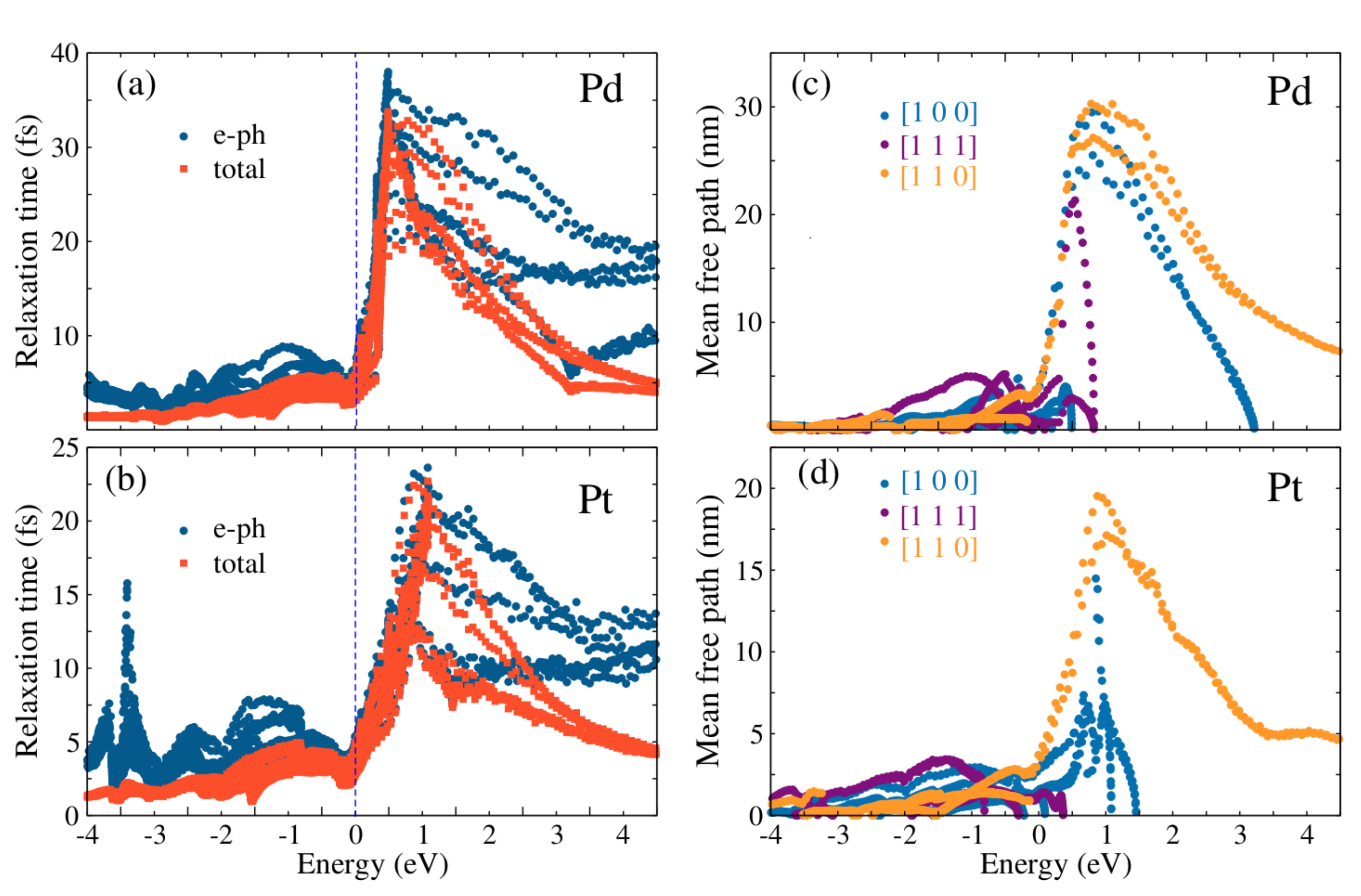}
\caption{\small{a) hot carrier relaxation time as a function of energy for Pd (top panel) and Pt (bottom panel). The scattering rates in green fully account for $e$-$e$ and $e$-$ph$ contribution, while the red curves only accounts for $e$-$ph$ contribution. The vertical blue dashed line indicates the Fermi level. b) Mean free path as a function of energy along the [100], [110], and [111] crystallographic directions.}}\label{fig3} 
\end{figure*}

Despite the experimental and theoretical works characterizing the carrier scattering rates of Pt group metals \cite{windmiller1969,ladstadter2004,gall2016}, the microscopic mechanism related to the electron-phonon relaxation rates is still poorly understood. 
Following an approach developed by Bernardi \emph{et al} in references \cite{bernardi2015,bernardi2014}, we employ a quantum mechanical microscopic approach based on \emph{ab initio} many-body perturbation theory to calculate the hot carrier scattering rates, taking into account both the $e$-$e$ and $e$-$ph$ contributions.
These are related to the imaginary part of the $e$-$e$ (Im$\Sigma^{e-e}_{n \textbf{k}}$) and $e$-$ph$ (Im$\Sigma^{e-ph}_{n \textbf{k}}$) self-energies. Figure \ref{fig2}-a shows the $e$-$e$ and $e$-$ph$ self-energy within a 5 eV energy window around $E_{F}$. For both metals, the $e$-$ph$ self-energy resembles the energy dependence of the electronic density of states (DOS). This is expected, since the DOS dictates the number of available electronic states that can effectively couple with lattice excitations. Clearly, the $e$-$ph$ contribution for Pd and Pt dominates within a narrow energy window of 0.8 eV around E$_F$, indicating that the energy loss within this range of energy is dominated by electron-phonon scattering processes. Concerning the $e$-$e$ self-energy, commonly related with the impact ionization rate, the results show that for Pd, the energy loss of hot carriers satisfying the condition $|E| > 4$ eV, is dominated by inelastic impact ionization. In contrast, for Pt, there is not a clear dominance of the $e$-$e$ interactions. Overall, the energy loss of hot holes is greater that of hot electrons by up to a factor of four. This feature has not been observed in gold group metals \cite{bernardi2015} within this energy range and can be ascribed to the proximity of $d$ states to the Fermi level \cite{bauer2015,zhukov2002}.
The $e$-$ph$ linewidth, $\Gamma^{e-ph}(E_{n \textbf{k}})=2$Im$\Sigma^{e-ph}_{n \textbf{k}}$, can also be used to estimate the electron-phonon mass enhancement parameter, $\lambda(E_{n \textbf{k}})$.
By considering, the $e$-$ph$ self-energy in the regime at which $k_{B}T$ is greater than the maximum phonon energy, one can write down a linear relation for the $e$-$ph$ linewidth with temperature \cite{grimvall1981}
\begin{equation}
\Gamma^{e-ph}(E_{n \textbf{k}}) = 2 \pi \lambda(E_{n \textbf{k}}) k_{B}T. \label{eq33}
\end{equation}
In order to estimate the electron-phonon mass enhancement parameter at the Fermi energy, $\lambda(E_{F})$, we computed the $e$-$ph$ linewidth, along the same high-symmetry directions used for the quasiparticle bandstructure (see Figure \ref{fig1}b) and then averaged over a narrow energy window of 10 meV around the $E_{F}$. 
The results, presented in the insets of Figure \ref{fig2}a, clearly show an increasing monotonic behavior of the linewidth with temperature. From equation (\ref{eq33}), we estimate the electron-phonon mass enhancement (the slope of the curves) to be $\lambda=0.66$ and $\lambda=0.78$ for Pd and Pt, respectively. Given that our calculations fully include the spin-orbit interaction, it can be argued that they partially take into account some effects related to the spin degree of freedom \cite{andersen1970}. This implies that the carrier lifetime at the Fermi level for Pt is greater than that of Pd (the greater $\lambda\left(E_{n\mathbf{k}}\right)$, the shorter the relaxation time). In this sense, the value obtained for Pd can be compared with the one estimated from specific heat measurements by Knapp and Jones \cite{knapp1972} ($\lambda=0.7$), Allen \cite{allen1987} ($\lambda=0.66$) and Savrasov \cite{savrasov1996} ($\lambda=0.69$), whose mass enhancement parameters include the electron-phonon and spin fluctuation contributions. Note that even though Pd and Pt present large mass enhancement parameters, superconductivity is absent in these metals due to the large paramagnon contribution \cite{hsiang1981}. The mass enhancement parameter obtained for Pt, on the other hand, over-estimates experimental measurements obtained from specific heat that span from 0.33 to 0.63 \cite{allen1987,knapp1972,andersen1970}. It is important to point out that the experimental results rely on \emph{ab initio} calculations for the density of states at the Fermi level. This parameter is, in fact, extremely sensitive towards the electronic bandstructure \cite{allen1987}, which at the time most of these studies were carried out, suffered from a computational limit on the number of \textbf{k}-grid samples, yielding less accurate values for the electronic structure.
Thus, we argue here, that this issue is related to a required theoretical input (the proper calculation of the density of states at Fermi level), which demands very fine \textbf{q}-and \textbf{k}-grid samples, as those employed in this work. In fact, more recent semi-empirical electron-phonon lifetimes estimated from resistivity measurements, reported by Wilson and Coh \cite{wilson2020} point out in the same direction. 

We proceed with the analysis of the electron-phonon coupling strength. Thus, in Figure \ref{fig2}b we present the calculated phonon dispersion at 0 K together with previous experimental neutron scattering measurements for Pd \cite{miiller1968} and Pt \cite{dutton1972}. We clearly observe a nice agreement between the compared curves. For completeness, in Figure \ref{fig2}c we present the phonon density of states in which one notes the presence of an intense peak at higher energies.
We then calculate the generalized Eliashberg spectral function, $\alpha^{2}F_{n\textbf{k}}(\omega)$, which determines the probability of specific phonon modes to decay into an electron-hole pair (see methods section). The Eliashberg spectral functions related to the highest occupied energy states of the high-symmetry points $\Gamma$, $X$, $L$ and $K$ (represented by green squares in Figure \ref{fig1}c) are shown in Figure \ref{fig2}d. In general, the selected spectral functions show considerable differences with the phonon density of states, especially in the intensity at low and high energies. This effect can be attributed to the complex Fermi surface that allows more effective Fermi nesting \cite{mueller1970,xu2020,andersen1970}.
In fact, Pd and Pt, present an intense peak around 16 meV and 13 meV, respectively, which are related to the contribution of transverse acoustic modes. The less intense peaks at higher frequencies of approximately 28 meV and 23 meV for, Pd and Pt respectively, is directly associated to the longitudinal acoustic mode. 

The hot carrier relaxation times, presented in Figure \ref{fig3}a-b, show that for Pd (Pt) hot electrons possess total relaxation times of up to 35 fs (24 fs), taking place at approximately 0.5 eV (1.0 eV) above the Fermi energy. In contrast, hot holes present faster relaxation times with a maximum of 6.8 and 5.6 fs for Pd and Pt, respectively, which are found close to the Fermi level. This result suggest that the extraction of hot holes before thermalization is a challenging task, since these carriers lose energy on the time scales below 5 fs.   
Note that the inclusion of $e$-$e$ scattering rates for Pt, drastically changes the relaxation times of hot holes for energies 1.0 eV below the Fermi level. A similar trend for hot electrons is observed at energies above 3 eV of the Fermi level. This results clearly highlight that a complete description of hot carriers in metals must include both the $e$-$e$ and $e$-$ph$ contributions. The carrier lifetimes at the Fermi level, calculated by averaging the relaxation time over a 10 meV energy window around the Fermi level, are 6.3 fs and 5.2 fs for Pd and Pt, respectively. These findings are in agreement with the electron-phonon lifetimes reported by Wilson and Coh \cite{wilson2020}, which employ semi-empirical models based on resistivity measurements yielding values of 8.6 fs for Pd and 6.1 fs for Pt.

Based on the results for the total carrier lifetime ($\tau_{n,\textbf{k}}$) and the carrier velocity $v_{n,\textbf{k}}=\frac{1}{\hbar}\frac{\partial E_{n,\textbf{k}}}{\partial \textbf{k}}$, obtained from the GW bandstructures, we calculate the mean free path, $\Lambda_{n,\textbf{k}}=\tau_{n,\textbf{k}}|v_{n,\textbf{k}}|$. Figure  \ref{fig3}c-d shows the carrier mean free path for hot holes and hot electrons along the $\Gamma X$, $\Gamma K$, and $\Gamma L$ directions of the Brillouin zone, which represent the [100], [110], and [111] crystallographic directions. As expected from the relaxation time results, hot holes in Pd an Pt possess mean free paths of less than 5 nm in all directions, making extraction challenging. In contrast, hot electrons in Pd (Pt), present mean free paths of up to 30 nm (20 nm). Such longer values are obtained for energies up to 0.8 eV above the Fermi level, and are ascribed to carrier velocities of hybridized $sp$ bands along the $\Gamma K$ direction, combined with the relative long lifetimes associated to the inherent low DOS of these bands. These results suggest that an efficient extraction of hot electrons in thin films of Pd (100) and Pd (110) may be achieved with excitation energies spanning from 0.4 to 1.6 eV. Accordingly, we predict that the efficient way of extracting hot electrons in Pt might be accomplished in thin films grown in the (110) direction for excitation energies within the 0.7 to 1.5 eV range. 
Our predictions are consistent with some available experimental measurements of optical absorption, electron-phonon mass enhancement and electron-phonon lifetimes. Nevertheless, this is not true for calculations in the absence of the spin-orbit coupling, which seems to understimate the optical absorption peaks while overstimates the carrier mean free path (see SI).
Finally, we should mention that our predicted scattering rates are in fact equilibrium rates based on standard \emph{ai}-MBPT, which are expected to be reliable at low carrier densities. These rates, however, can be employed as inputs in a complete description of carrier dynamics based on the evolution of the non-equilibrium Green's functions, which has been recently devised and employed to describe the real-time carrier dynamics in some paradigmatic materials \cite{marini2012,sangalli2015,molina2017}. The latter is, however, beyond of the scope of this work.
\section{Conclusions}
 Based on \emph{ab initio} many body perturbation theory, we provided a comprehensive theoretical description of the hot carrier generation, carrier lifetimes, and carrier transport in bulk Pd and Pt.
 Our optical response results showed that the photoexcited carriers in these metals are mainly generated in the near-infrared range (from 0.3 to 0.78 eV) by direct optical transitions, while intraband transitions are relevant at infrared frequencies. We also found that the electron-phonon mass enhancement parameter for Pt is slightly higher than Pd, a striking result that shed light into the understanding of previous experimental measurements. Our findings indicate that the extraction of hot holes in Pt and Pd is a challenging task as these carriers possess sub-5-fs lifetimes. In addition, we showed that thin films of Pd (110) and Pd (100) subject to excitation energies ranging from 0.4 to 1.6 eV can be used for efficient hot electron generation and extraction as a consequence of its 30 nm mean free path, comparable to that of noble metals.
\section{Acknowledgements}
A.R.R. acknowledges support from ICTP-SAIRF (FAPESP project 2011/11973-4). M.S.L. acknowledges the financial support from the National Science Foundation (award number 20-16617). A.M. acknowledges the funding received from the European Union projects: MaX {\em Materials design at the eXascale} H2020-EINFRA-2015-1, Grant agreement n. 676598, and H2020-INFRAEDI-2018-2020/H2020-INFRAEDI-2018-1, Grant agreement n. 824143;  {\em Nanoscience Foundries and Fine Analysis - Europe} 
H2020-INFRAIA-2014-2015, Grant agreement n. 654360. This work uses the computational resources from GRID-UNESP.

\section{APPENDIX}
\subsection{Theory and computational details}
We consider pristine face centered cubic crystals of Pd and Pt with fully relaxed lattice parameters 
of 3.86 and 3.91 \AA, respectively.
Our calculations are conducted in two steps. First, plane-wave density functional theory is used to obtain the electronic ground-state within the local density approximation (LDA) for the exchange-correlation potential. We employed norm-conserving pseudopotentials treating the semicore $s$ and $p$ states as valence electrons, a 90 Ry kinetic energy cutoff for Pd/Pt and a $k$-sampling grid in the Monkhorst-Pack scheme of 12$\times$12$\times$12 as implemented in the Quantum-Espresso code \cite{pwscf}. The structures were fully optimized to their equilibrium position with pressures on the lattice unit cell smaller than 0.02 kbar. The Marzari-Vanderbilt smearing scheme including a width of 0.02 Ry is employed. 
Next, we use density functional perturbation theory (DFPT) \cite{dfpt} to compute the vibrational frequencies $\omega_{\textbf{q}\lambda}$ and the derivatives of the self-consistent Khon-Sham potential with respect to the atomic displacements, inputs required  to evaluate the electron-phonon coupling matrix elements, $g^{\textbf{q}\lambda}_{nn'\textbf{k}}$, which represent the probability amplitude for an electron to be scattered due to emission or absorption of phonons,
\begin{equation}
\begin{split}
g^{\textbf{q}\nu}_{nn'\textbf{k}}=\sum_{s \alpha}[2M_{s}\omega_{\textbf{q} \nu}]^{-1/2}e^{i\textbf{q}.\tau_{s}} \xi_{\alpha}(\textbf{q} \nu|s) \\
\times \langle n'\textbf{k-q}|\frac{\partial V_{scf}(\textbf{r})}{\partial R_{s \alpha}}|n\textbf{k} \rangle, \label{eq5.1}
\end{split}
\end{equation}
where, $M_{s}$ is the atomic mass of the s$-$th atom, $\tau_{s}$ is the position of the atomic displacement in the unit cell,
$\xi_{\alpha}(\textbf{q}\nu|s)$ are the components of the phonon polarization vectors, and $V_{scf}(\textbf{r})$ is the self-consistent DFT ionic potential. 

Many-Body perturbation theory (MBPT) \cite{Onida2002} is used to describe the electron-phonon interaction which is treated perturbatively \cite{marini2015} by considering the first order Taylor expansion term in the nuclear displacement, commonly known as the Fan term. Given that the Fan self-energy term is a complex function, it is responsible for providing the imaginary part of the quasiparticle energies that correspond to the quasiparticle linewidths. Thus, the Fan self-energy is given by
\begin{equation}
 \begin{split}
 \Sigma^{Fan}_{n\textbf{k}}(\omega,T)=\sum_{n'\textbf{q} \nu}\frac{|g^{\textbf{q}\nu}_{nn'\textbf{k}}|^2}{N_{q}} \Big[\frac{N_{
 \textbf{q}\nu}(T)+1-f_{n'\textbf{k-q}}}{\omega-\epsilon_{n'\textbf{k-q}}-\omega_{\textbf{q}\nu}-i\delta} \\
+\frac{N_{\textbf{q}\nu}(T)+f_{n'\textbf{k-q}}}{\omega-\epsilon_{n'\textbf{k-q}}+\omega_{\textbf{q}\nu}-i\delta}\Big], \label{eq4}
 \end{split}
\end{equation}
where $N_{\textbf{q}\lambda}$ and $f_{n'\textbf{k-q}}$ represent the Bose-Einstein and Fermi-Dirac distribution functions, while N$_{q}$ is the number of q vectors in the Brillouin zone. The electron-phonon relaxation time is given by,
\begin{equation}
 \tau_{n,\textbf{k}} = \frac{\hbar}{2} Im [\Sigma^{Fan}_{n\textbf{k}}]^{-1}.
\end{equation}

We employ both, the EPW \cite{ponce2016} and YAMBO \cite{sangalli2019} code to compute the Fan self-energy and the generalized Eliashberg spectral functions, respectively. The electron-phonon matrix elements required for the EPW code are initially calculated on a course 4$\times$4$\times$4 \textbf{q}-grid, and then interpolated onto fine grids using maximally localized Wannier functions (MLWFs) \cite{mostofi2016}. The random projection method, including eighteen (nine) Wannier functions in the presence (absence) of spin-orbit coupling, was used to construct MLWFs.

We use a small Lorentzian broadening of 10 meV and a fine randomly distributed $\textbf{q}$-vectors mesh to better map out the phonon transferred momentum \cite{q-random}. Thus, the electron phonon self-energy calculation obtained with EPW (YAMBO) uses a 40$\times$40$\times$40 (26$\times$26$\times$26) \textbf{k}-grid and 2.7$\times$10$^{4}$ (5$\times$10$^{2}$) random \textbf{q}-points in the Brillouin zone. We found that the converged parameters for both codes provide similar results for the e-ph self-energy (See supplementary information).

The generalized Eliashberg spectral function is be obtained by the relation
\begin{equation}
\alpha^{2}F_{n\textbf{k}}(\omega)=\frac{1}{N}\sum_{n^{'}\textbf{q}\lambda }\Big[\frac{|g^{\textbf{q}\lambda}_{nn'\textbf{k}}|^2}{\epsilon_{n\textbf{k}}
-\epsilon_{n'\textbf{k-q}}}\Big] \times \delta(\omega-\omega_{\textbf{q}\lambda}), \label{eq3}
\end{equation}
which enables us to visualize the e-ph coupling strength for a given state $|n\textbf{k} \rangle$.

The real and imaginary part of the electron-electron self-energy (Im$\Sigma^{e-e}_{n \textbf{k}}$) is computed within the full frequency GW approximation as implemented in the YAMBO code. An energy cutoff of 68 Ry for the exchange potential of Pd/Pt, respectively. The energy cutoff for the polarization function is set at 20 Ry for Pd/Pt. For both metals, a \textbf{q}-vectors grid of 18 $\times$ 18 $\times$ 18 together with 240 unoccupied bands were employed for the polarization and Green's function. The frequency dependent dielectric function is computed over 400 frequencies along the real axis for its numerical evaluation.
Finally, The optical absorption is calculated at the Random Phase Approximation level in a grid of 56 $\times$ 56 $\times$ 56, taking into account the electronic quasiparticle corrections computed within full frequency GW approximation.
%

\end{document}